\documentclass[a4paper,11pt]{article}
\pdfoutput=1 % if your are submitting a pdflatex (i.e. if you have
\usepackage{jheppub} % for details on the use of the package, please
\usepackage[T1]{fontenc} % if needed
\usepackage{multirow}
\usepackage{subcaption}
\usepackage{tikz}

\usepackage{pdflscape}

\usetikzlibrary{matrix, positioning}

\def\la{\langle}
\def\ra{\rangle}

\def\spab#1#2{\la #1 \vert \gamma^\mu \vert #2 ]}
\def\spba#1#2{[ #1 \vert \gamma^\mu \vert #2 \ra}

\title{Top quark contribution to two-loop helicity amplitudes for \boldmath{Z} boson pair production in gluon fusion}

\author{Christian Br\o{}nnum-Hansen}
\author{and Chen-Yu Wang}

\affiliation{Institute for Theoretical Particle Physics, KIT, Karlsruhe, Germany}

\emailAdd{christian.broennum-hansen@kit.edu}
\emailAdd{chen-yu.wang@kit.edu}

\abstract{We compute the top quark contribution to the two-loop amplitude for on-shell $Z$ boson pair production in gluon fusion, $gg \to ZZ$. Exact dependence on the top quark mass is retained.
For each phase space point the integral reduction is performed numerically and the master integrals are evaluated using the auxiliary mass flow method, allowing fast computation of the amplitude with very high precision.}

\keywords{Perturbative QCD, Scattering Amplitudes}
\preprint{TTP21-002, P3H-21-008}

\begin{document}
\maketitle
\flushbottom

\section{Introduction}

Production of $Z$ boson pairs is an important process at the LHC.
The gluon fusion channel, $gg \to ZZ$, is loop-induced. For this reason, it is suppressed by the strong coupling constant $\alpha_{s}$ in comparison to the quark annihilation channel $q \overline{q} \to ZZ$ which enters at tree level.
However, the large gluon flux as well as event selection enhance the contribution of the gluon fusion channel to the hadronic cross section~\cite{Binoth:2006mf}.
Therefore this production mode is essential for a reliable description of $Z$ boson pair production.

The current status of the amplitude calculations for this process is as follows.
The one-loop amplitude was calculated long ago~\cite{Glover:1988rg,Glover:1988fe}.
The two-loop amplitude for massless internal quarks is also known~\cite{Caola:2015ila,vonManteuffel:2015msa}.
However, until very recently, contributions of massive quarks have only been calculated approximately~\cite{Melnikov:2015laa,Davies:2020lpf}.
The goal of this paper is to present a calculation of the $gg \to ZZ$ two-loop amplitude keeping the dependence on the top quark mass.
We note that when this paper was being completed, ref.~\cite{Agarwal:2020dye} appeared where the two-loop amplitude with full top quark mass effects was calculated using analytic integral reduction and numerical integration using sector decomposition~\cite{Binoth:2000ps,Bogner:2007cr}.

To compute the $gg \to ZZ$ amplitude, we largely follow the method described in our paper on a similar process, $gg \to WW$~\cite{Bronnum-Hansen:2020mzk}.
The main difference lies in the integral reduction.
For the present calculation, we decided to perform integration-by-part reductions individually for each phase space point.
As before, the master integrals are evaluated efficiently using a system of ordinary differential equations and the auxiliary mass flow method~\cite{Liu:2017jxz,Liu:2020kpc}.

The organisation of this paper is as follows.
In Section~\ref{sec:calculation} we describe the amplitude calculation, discuss the ultraviolet and infrared divergences that arise and summarise the renormalisation procedure.
In Section~\ref{sec:numericalevaluation} we discuss the evaluation of the master integrals.
A benchmark point for the amplitude for physical kinematics and plots for the helicity amplitudes across partonic phase space are presented in Section~\ref{sec:helicityamplitudes}.
We conclude in Section~\ref{sec:conclusions}.

\section{Calculational setup}\label{sec:calculation}

We consider the two-loop amplitude for $Z$ boson production in gluon fusion
\begin{align}
    \label{equ:process}
    g(p_1) + g(p_2) \to Z(p_3) + Z(p_4)
    \text{.}
\end{align}
This process is mediated by quark loops.
We calculate the contribution where the external $Z$ bosons couple directly to top quarks and disregard other quark flavours.
The exact dependence on the quark mass $m_t$ is retained.
We do not consider diagrams with an intermediate $\gamma$-, $Z$- or Higgs-boson.
The photon mediated amplitude vanishes identically and calculations for the two latter are available in the literature~\cite{Kniehl:1989qu,Gounaris:2000tb,Spira:1995rr,Harlander:2005rq,Anastasiou:2006hc}.
All external particles are on-shell
\begin{align}
p_{1}^{2} = p_{2}^{2} = 0,\quad p_{3}^{2} = p_{4}^{2} = m_{Z}^{2}\text{.}
\end{align}
We define the usual Mandelstam variables
\begin{align}
    s = (p_{1} + p_{2})^{2},\quad
    t = (p_{1} - p_{3})^{2},\quad
    u = (p_{2} - p_{3})^{2},
\end{align}
which satisfy the relation $s + t + u = 2 m_Z^2$.

For four-dimensional external states the amplitude $A$ can be decomposed in terms of 18 tensor structures
\begin{align}
    A(\{p_{i}\}, \{\epsilon_{j}\}, m_{t})
    & =
    \sum_{I = 1}^{18} A_{I}(s, t, m_{W}, m_{t})\ T_{I}^{\mu \nu} (\{p_{i}\}, \{\epsilon_{j}\}) \epsilon_{3 \mu}^{*}(p_{3}) \epsilon_{4 \nu}^{*}(p_{4}). \label{eq:ampdecomp}
\end{align}
Their definitions are given in ref.~\cite{Binoth:2006mf} and reproduced in appendix~\ref{sec:tensors} for convenience. Our goal is to calculate the form factors $A_{I = 1, ..., 18}$.

\subsection{Pole structure}\label{sec:polestructure}

The form factors in eq.~\eqref{eq:ampdecomp} contain ultraviolet (UV) and infrared (IR) divergences.
In order to regulate them we employ dimensional regularisation.
We work in space-time with dimension $d = 4 - 2 \epsilon$.

Expanding the unrenormalised amplitude, $\widehat{A}$, in the bare strong coupling constant $\widehat{\alpha}_{s}$ we write
\begin{align}
\widehat{A} = \frac{\widehat{\alpha}_{s}}{2 \pi} \widehat{A}^{(1)} + \left( \frac{\widehat{\alpha}_{s}}{2 \pi} \right)^{2} \widehat{A}^{(2)} + \mathcal{O}\left( \widehat{\alpha}_{s}^{3} \right).
\end{align}
We employ the same renormalisation scheme as used in ref.~\cite{Bronnum-Hansen:2020mzk} where the gluon field $G_{\mu}$ and the top quark mass $m_{t}$ are computed in the on-shell scheme while the strong coupling constant, $\alpha_{s}$, is renormalised in the $\overline{\text{MS}}$ scheme.
The relations between bare and renormalised quantities read
\begin{align}
\label{equ:renormalisationconstants}
\widehat{\alpha}_{s} = \mu^{2 \epsilon} S_{\epsilon} Z_{\alpha_{s}} \alpha_{s},
\qquad
\widehat{G}_{\mu} = \sqrt{Z_{g}} G_{\mu},
\qquad
\widehat{m}_{t} = Z_{m_{t}} m_{t},
\end{align}
where $\mu$ is the renormalisation scale and $S_{\epsilon} = (4 \pi)^{- \epsilon} e^{\epsilon \gamma_{E}}$.
All renormalisation constants are expanded in the strong coupling
\begin{align}
Z = \sum_{n = 0} \left( \frac{\alpha_{s}}{2 \pi} \right)^{n} Z^{(n)},
\qquad
Z^{(0)} = 1.
\end{align}
The renormalised amplitude is related to the unrenormalised one by
\begin{align}
\label{equ:renormalisedamplitude}
A (\epsilon, \mu, \alpha_{s}, m_{t})
& =
Z_{g} \widehat{A} (\epsilon, \widehat{\alpha}_{s}, \widehat{m}_{t})
=
\frac{\alpha_{s}}{2 \pi}
A^{(1)}(\epsilon, m_{t})
+
\left( \frac{\alpha_{s}}{2 \pi} \right)^{2}
A^{(2)}(\epsilon, m_{t})
+
\mathcal{O}(\alpha_{s}^{3}),
\\
A^{(1)}(\epsilon, m_{t})
& =
\mu^{2 \epsilon} S_{\epsilon}
\widehat{A}^{(1)} (\epsilon, m_{t}),
\\
A^{(2)}(\epsilon, m_{t})
& =
\mu^{2 \epsilon} S_{\epsilon}
\left[
(Z_{g}^{(1)} + Z_{\alpha_{s}}^{(1)})
\widehat{A}^{(1)} (\epsilon, m_{t})
+
m_{t} Z^{(1)}_{m_{t}}
\widehat{C}^{(1)} (\epsilon, m_{t})
\right]
\nonumber
\\
& \phantom{= {}}
+
\left( \mu^{2 \epsilon} S_{\epsilon} \right)^{2}
\widehat{A}^{(2)} (\epsilon, m_{t}).\label{eq:A2ren}
\end{align}
The mass counterterm, $\widehat{C}^{(1)}$, is calculated as a separate amplitude.

The relevant renormalisation factors read~\cite{Gross:1973id, Politzer:1973fx, Melnikov:2000zc, Beenakker:2002nc, Czakon:2007wk}
\begin{align}
\label{equ:alphas}
Z_{\alpha_s}^{(1)}
& =
- \frac{\gamma_{g}}{\epsilon},
\\
Z_{g}^{(1)}
& =
S_{\epsilon}
\left( \frac{4 \pi \mu^{2}}{m_{t}^{2}} \right)^{\epsilon}
\Gamma(1 + \epsilon)\,
T_{F}
\left[ - \frac{2}{3 \epsilon} \right]
,\label{eq:Zg}
\\
Z_{m_{t}}^{(1)}
& =
S_{\epsilon}
\left( \frac{4 \pi \mu^{2}}{m_{t}^{2}} \right)^{\epsilon}
\Gamma(1 + \epsilon)\,
C_{F}
\left[ - \frac{3}{2 \epsilon} - \frac{2}{1 - 2 \epsilon} \right]
,
\end{align}
where $\gamma_{g} = \frac{11}{6} C_{A} - \frac{2}{3} T_{F} (n_{l} + 1)$, $n_l$ is the number of massless quarks, and $T_{F} = \frac{1}{2}$.

After renormalisation the remaining poles are of IR origin. Their structure is predicted by the Catani formula~\cite{Catani:1998bh}.
Since $gg \to ZZ$ is loop-induced and has trivial colour structure, the IR poles are particularly simple.
By subtracting the IR poles we define a finite remainder,
\begin{align}
\label{eq:irfactorisation}
F^{(2)}(\epsilon, \mu)
= A^{(2)}(\epsilon, \mu) - I^{(1)} (\epsilon, \mu) A^{(1)}(\epsilon, \mu).
\end{align}
In the present case the Catani operator reads
\begin{align}
\label{equ:iroperator}
I^{(1)} (\epsilon, \mu)
& =
- \frac{e^{\epsilon \gamma_{E}}}{\Gamma (1 - \epsilon)}
\left( \frac{C_{A}}{\epsilon^{2}} + \frac{\widetilde{\gamma}_{g}}{\epsilon} \right)
\left( \frac{\mu^{2} e^{i \pi}}{s} \right)^{\epsilon},
\end{align}
where $\widetilde{\gamma}_{g} = \frac{11}{6} C_{A} - \frac{2}{3} T_{F} n_{l}$. The $\epsilon$-dependence of the leading order amplitude $A^{(1)}(\epsilon, \mu)$ is important for computing the finite remainder.

From here on we disregard the trivial contribution proportional to $n_l$, which in our calculation enters only through eqs.~\eqref{equ:alphas} and~\eqref{equ:iroperator} and contributes a finite term proportional to the leading order amplitude.

\subsection{Amplitude calculation and integral reduction}\label{sec:amplitude}

Using \texttt{QGRAF}~\cite{Nogueira:1991ex} we generate 8 and 138 diagrams for the one- and two-loop amplitudes respectively.
Colour and Dirac algebra is performed in \texttt{FORM}~\cite{Vermaseren:2000nd,Kuipers:2013pba,Ruijl:2017dtg}. Symmetries between diagrams are established using \texttt{REDUZE 2}~\cite{vonManteuffel:2012np} and we find that all two-loop diagrams can be mapped on to 21 integral families.
The two-loop diagrams are classified according to the colour factors $C_A$ and $C_F$.
Factorisable two-loop diagrams are present in the $t$- and $u$ channels only.
A total of 37 two-loop diagrams vanish due to colour conservation, leaving 101 non-vanishing diagrams.
Representative diagrams and their colour classifications are given in Table~\ref{tab:colour}.

\begin{table}[ht]
	\centering
	\begin{tabular}{|c|c|c|}
		\hline \hline
		& \textbf{Colour factor} & \textbf{\# of non-vanishing diagrams} \\
		\hline \hline
		\multirow{6}{*}{\includegraphics[width=0.15\linewidth]{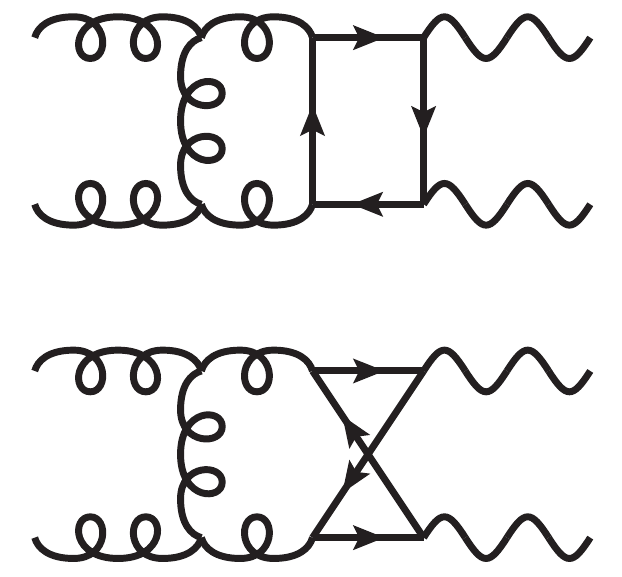}} & & \\
		& & \\
		& $C_A$ & 33 \\
		& & \\
		& & \\
		& & \\
		\hline
		\multirow{3}{*}{\includegraphics[width=0.15\linewidth]{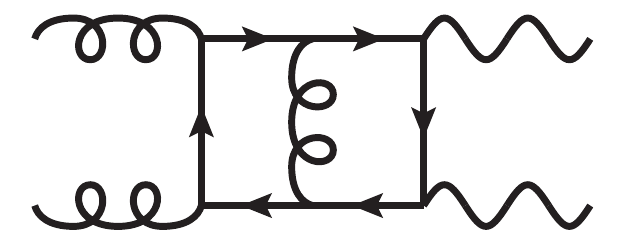}} & & \\
		& $C_F$ & 40 \\
		& & \\
		\hline
		\multirow{3}{*}{\includegraphics[width=0.15\linewidth]{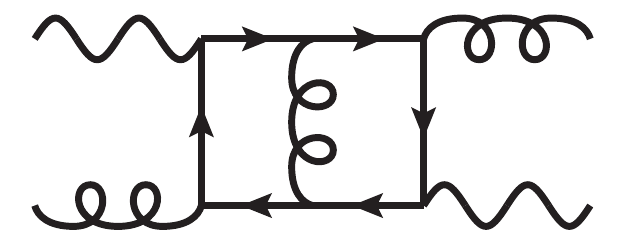}} & & \\
		& $C_A - 2 C_F$ & 20 \\
		& & \\
		\hline
		\multirow{3}{*}{\includegraphics[width=0.15\linewidth]{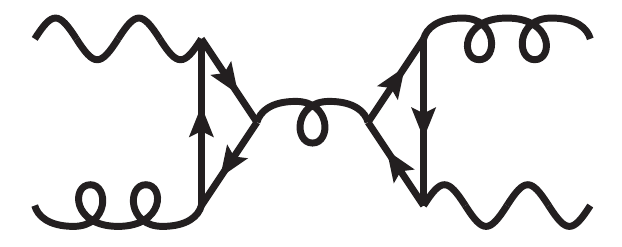}} & & \\
		& $1$ & 8 \\
		& & \\
		\hline \hline
	\end{tabular}
	\caption{The classification of representative diagrams by colour factors. Curly lines represent gluons, wavy lines are $Z$ bosons and straight lines represent top quarks. All nonplanar diagrams come with colour factor $C_A$.}
	\label{tab:colour}
\end{table}

We decompose the two-loop amplitude according to the colour factors
\begin{align}
    \mathcal{A}
    & =
    \delta^{c_1 c_2} A,
    \\
    A & =
    C_A A^{[C_A]} + C_F A^{[C_F]} + A^{[\triangle^{2}]},
\end{align}
where $c_{1,2}$ are colour indices of the incoming gluons.
From the discussion of the pole structure we note that $A^{[\triangle^{2}]}$, which is formed entirely of the factorisable diagrams, is finite.\footnote{After UV renormalisation, the part of the two-loop amplitude with unity colour factor receives a finite contribution from $Z_{g}$ defined in eq.~\eqref{eq:Zg}.}
$A^{[C_F]}$ is finite after UV renormalisation and hence IR poles are only found in $A^{[C_A]}$.

The $Z$ bosons couple to fermions through vector and axial currents.
The $gg \to ZZ$ process involves two such vertices and can therefore be split into three parts; vector-vector, vector-axial, and axial-axial.
The vector-axial part vanishes due to charge parity conservation and the two remaining parts can be considered separately. For each colour factor we have
\begin{align}
A^{[X]} = g_V^2 \left(A^{[X]}_{vv} + \frac{g_A^2}{g_V^2} A^{[X]}_{aa}\right),\label{eq:vvaadecomp}
\end{align}
where $X = C_A,\ C_F,\ \triangle^{2}$.
The weak couplings are given by
\begin{align}
g_V = \frac{e}{2 \sin (2 \theta_W)} \left(1 - \frac{8}{3} \sin ^2 (\theta_W) \right),\qquad g_A = \frac{e}{2 \sin (2 \theta_W)},
\end{align}
where $\sin ^2 \theta_W = 1 - m_W^2 / m_Z^2$ is the weak mixing angle and $e = \sqrt{4 \pi \alpha}$ is the absolute value of the electron charge. We note that the vector-vector part of the factorisable diagrams, $A^{[\triangle^{2}]}_{vv}$, vanishes due to Furry's theorem.

In all non-factorisable diagrams both $Z$ bosons are attached to the same quark loop leading to a single Dirac trace involving zero or two $\gamma_5$ matrices. These single traces can be evaluated naively in $d$ dimensions by using the anti-commutative property of $\gamma_5$.
The factorisable diagrams contain traces involving a single $\gamma_5$. For these diagrams we employ the Larin scheme~\cite{Larin:1993tq}, replacing the axial current by
\begin{align}
\gamma_{\mu} \gamma_{5} = \frac{i}{3 !} \varepsilon_{\mu \nu \rho \sigma} \gamma^{\nu} \gamma^{\rho} \gamma^{\sigma}.
\end{align}
As this class of diagrams consists of products of one-loop anomaly diagrams, no finite renormalisation is required.

After projection onto the tensor structures introduced in eq.~\eqref{eq:ampdecomp}, the form factors, $A_{I}$, are in a form suitable for integration-by-parts (IBP) reduction.
At this point we switch to an entirely numerical approach in which masses are chosen to be numbers close to their experimental values,\footnote{The integer values chosen for the particle masses are within a few per mille of the current best estimates~\cite{Zyla:2020zbs}. This approximation should have a negligible effect on phenomenological applications.}
\begin{align}
    m_{t} = 173 \text{~GeV},
    \quad
    m_{Z} = 91 \text{~GeV}.\label{eq:fixedmasses}
\end{align}
We perform numerical reductions to master integrals across partonic phase space using rational values for $s$ and $t$.
Having space-time dimension $d$ as the only free parameter keeps all intermediate expressions compact and manageable.

In addition to the computational advantage of working numerically, we also carefully choose the master integral basis to limit the size of the reduction tables and computation time.
Even in the purely numerical case, significant simplification is achieved by avoiding reduction coefficients with denominators that are not factorised in kinematic invariants and space-time dimension.
We find that by introducing numerator insertions and increasing denominator powers, non-factorisable denominators are eliminated.

The reductions to master integals required for the evaluation of the amplitude can be performed for a single phase space point in approximately 3 hours on a single CPU core using \texttt{KIRA} 2.0~\cite{Klappert:2020nbg}; speed strongly depends on fast read/write access to disk drives.
Each reduction requires between 3 and 4 gigabytes of memory.
The short runtime and low memory consumption enable straightforward parallelisation for a large number of phase space points.

The approach to integral reduction described here is in variance to our approach in previous work~\cite{Bronnum-Hansen:2020mzk} where reductions were performed with $s$ and $t$ as free parameters.
We kept the reductions tractable by projecting integrals onto one master integral at a time which, however, in turn made bookkeeping more involved.
We find the strategy adopted in the present calculation with straightforward parallelisation and compact univariate reduction tables to be more efficient.

\section{Numerical evaluation}\label{sec:numericalevaluation}

\begin{figure}[ht]
    \centering
    \begin{subfigure}[ht]{0.3\linewidth}
        \centering
        \includegraphics[width=0.8\linewidth]{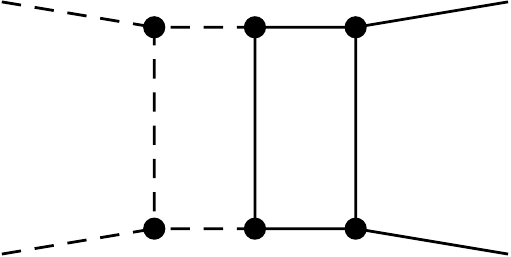}
        \caption{planar no.\ 1}
    \end{subfigure}
    ~
    \begin{subfigure}[ht]{0.3\linewidth}
        \centering
        \includegraphics[width=0.8\linewidth]{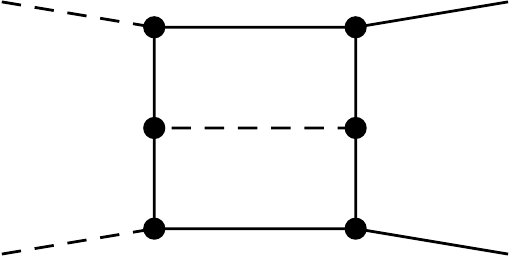}
        \caption{planar no.\ 2}
    \end{subfigure}
    ~
    \begin{subfigure}[ht]{0.3\linewidth}
        \centering
        \includegraphics[width=0.8\linewidth]{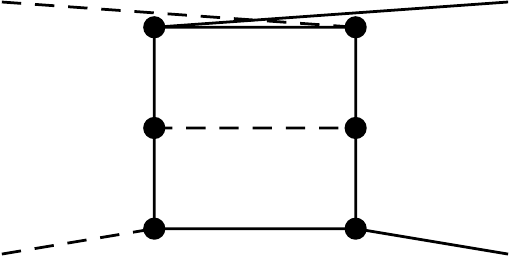}
        \caption{planar no.\ 3}
    \end{subfigure}
    \\
    \begin{subfigure}[ht]{0.3\linewidth}
        \centering
        \includegraphics[width=0.8\linewidth]{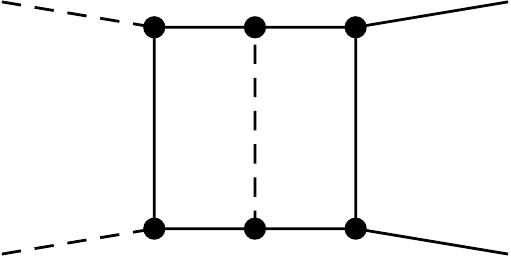}
        \caption{planar no.\ 4}
    \end{subfigure}
    ~
    \begin{subfigure}[ht]{0.3\linewidth}
        \centering
        \includegraphics[width=0.8\linewidth]{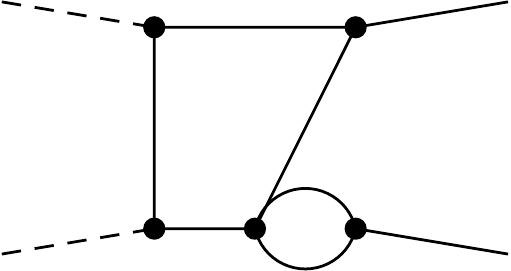}
        \caption{planar no.\ 5}
    \end{subfigure}
    ~
    \begin{subfigure}[ht]{0.3\linewidth}
        \centering
        \includegraphics[width=0.8\linewidth]{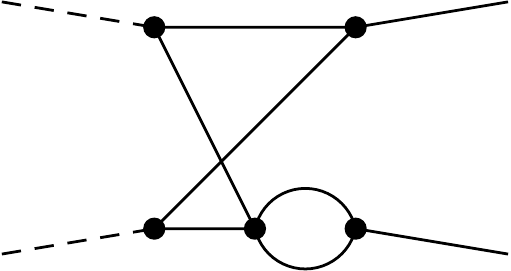}
        \caption{planar no.\ 6}
    \end{subfigure}
    \\
    \begin{subfigure}[ht]{0.3\linewidth}
        \centering
        \includegraphics[width=0.8\linewidth]{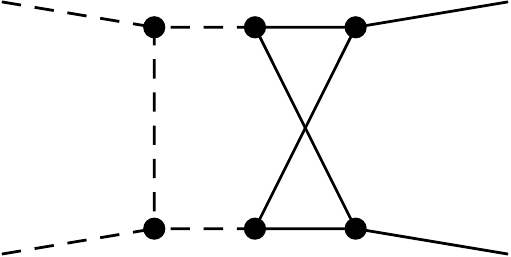}
        \caption{nonplanar no.\ 1}
    \end{subfigure}
    ~
    \begin{subfigure}[ht]{0.3\linewidth}
        \centering
        \includegraphics[width=0.8\linewidth]{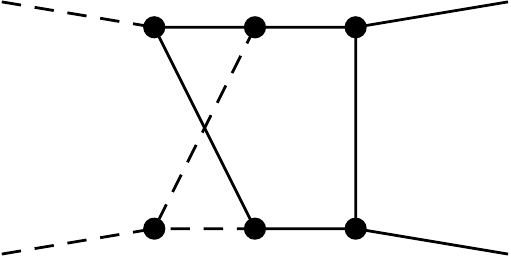}
        \caption{nonplanar no.\ 2}
    \end{subfigure}
    ~
    \begin{subfigure}[ht]{0.3\linewidth}
        \centering
        \includegraphics[width=0.8\linewidth]{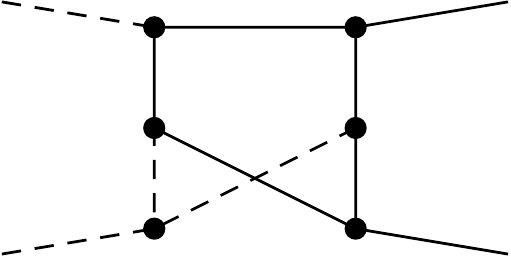}
        \caption{nonplanar no.\ 3}
    \end{subfigure}
    \caption{Topologies of integral families.
        Solid and dashed lines correspond to massive and massless particles respectively.
        Internal massive particles have mass $m_{t}$ while external massive particles have mass $m_{Z}$.
        The first 5 planar topologies and nonplanar no.\ 3 can be crossed ($p_{1} \leftrightarrow p_{2}$) giving a total of 15 topologies.}
    \label{fig:family}
\end{figure}

Having expressed the full amplitude through master integrals,
we need to evaluate them.
The master integrals are defined as follows
\begin{equation}
    I(a_{1}, \ldots, a_{9})
    =
    \int \left( \prod_{n = 1}^{2} e^{\epsilon \gamma_{E}} \frac{\mathrm{d}^{d} l_{n}}{i \pi^{d / 2}} \right)
    \frac{1}{D_{1}^{a_{1}} D_{2}^{a_{2}} \cdots D_{9}^{a_{9}}}
    \text{,}
\end{equation}
where $D_{i}$ are denominators that appear in one of the 15 families given in appendix~\ref{sec:family}.
Their topologies are shown in figure~\ref{fig:family}.
Note that we absorb a factor of $- i (4 \pi)^{2 - \epsilon} e^{\epsilon \gamma_{E}}$ per loop into the definition of the master integrals.

To compute the 205 two-loop master integrals we follow the same procedure as described in ref.~\cite{Bronnum-Hansen:2020mzk} and employ the auxiliary mass flow method~\cite{Liu:2017jxz,Liu:2020kpc}.
We construct a system of differential equations with respect to $m_{t}^{2}$
and solve it starting from the boundary conditions at $m_{t}^{2} \to -i \infty$ and moving to the physical value $m_{t} = 173 \text{~GeV}$.\footnote{We need to keep $m_{t}$ as a parameter when constructing the differential equation using IBP reduction, but this introduces no difficulty in this problem.}
Evaluating all 205 master integrals to 20 digits at a typical phase space point takes less than 1 hour on a single CPU core.

\begin{figure}[ht]
    \centering
    \begin{tikzpicture}
        \matrix (mat) [left delimiter=\lbrace]
        {
            \node {region 1:}; &
            \node (r1a) {$m_{t}^{-6 - 4 \epsilon} \times$};
            \node (r1b) [right=0 of r1a] {\includegraphics[width=0.2\linewidth]{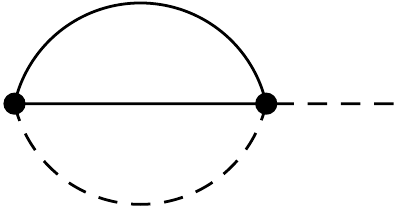}};
            \\
            \node {region 2:}; &
            \node (r2a) {$m_{t}^{-4 - 2 \epsilon} \times$};
            \node (r2b) [right=0 of r2a] {\includegraphics[width=0.2\linewidth]{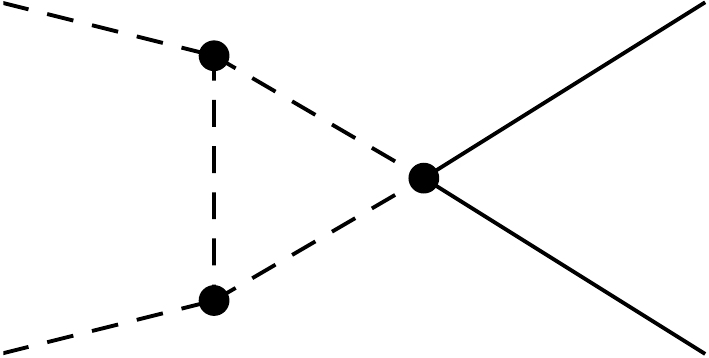}};
            \node (r2c) [right=0 of r2b] {$\times$};
            \node (r2d) [right=0 of r2c] {\includegraphics[width=0.2\linewidth]{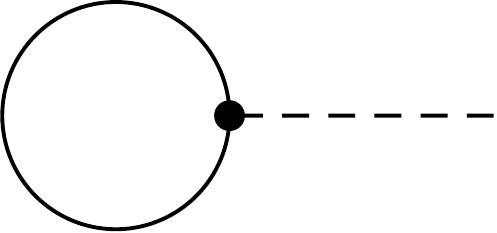}};
            \\
        };
        \node (propto) [left=15pt of mat] {$\propto$};
        \node (diagram) [left=5pt of propto] {\includegraphics[width=0.2\linewidth]{figures/family_planar1.pdf}};
    \end{tikzpicture}
    \caption{A typical master integral and its leading regions.
        Solid and dashed lines describe massive and massless particles respectively.
        All internal massive particles have masses equal to $m_{t}$, while all external massive particles have masses equal to $m_{Z}$.}
    \label{fig:region}
\end{figure}

\begin{figure}[ht]
    \centering
    \begin{subfigure}[ht]{0.3\linewidth}
        \centering
        \includegraphics[width=0.8\linewidth]{figures/boundary_i_1.pdf}
        \caption{$I_{1}$}
    \end{subfigure}
    ~
    \begin{subfigure}[ht]{0.3\linewidth}
        \centering
        \includegraphics[width=0.8\linewidth]{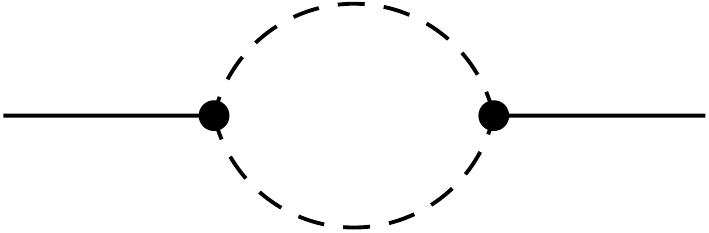}
        \caption{$I_{2}$}
    \end{subfigure}
    \caption{Master integrals for the boundary conditions.
        Solid and dashed lines correspond to massive and massless particles respectively.
        See appendix~\ref{sec:boundary} for their explicit expressions~\cite{tHooft:1978jhc}.
        }
    \label{fig:boundary}
\end{figure}

Comparing to the calculation of the $g g \to W W$ amplitude, we have more massive propagators in the $g g \to Z Z$ case,
which leads to fewer regions at the boundary and simpler boundary conditions.
Indeed, we find that the computation of the boundary conditions only requires 2 regions:
\begin{enumerate}
    \item All internal momenta are comparable to $m_{t}^{2} \to -i \infty$.
    \item Some internal momenta that form a closed loop are comparable to $m_{t}^{2} \to -i \infty$,
        while the remaining momenta are much smaller than $m_{t}^{2}$
        and are comparable to other kinematic parameters (i.e.\ $s$, $t$, $m_{Z}^{2}$).
\end{enumerate}
In figure~\ref{fig:region} we show a typical master integral in these two regions.
In each region we compute the large mass expansion of the integral.
The expansion coefficients can be expressed in terms of the original integral with some of the propagators contracted,
together with an overall power of $m_{t}$.
All of these contracted integrals can in turn be expressed by the two one-loop integrals listed in figure~\ref{fig:boundary} through IBP reduction.

We cross-check the evaluation of the master integrals against \texttt{pySecDec}~\cite{Borowka:2017idc,Borowka:2018goh}
at a physical phase space point above the top quark threshold
\begin{equation}
    \label{equ:crosscheck_point}
    s = \frac{400}{19} \times (91 \text{~GeV})^{2}
    \text{,}
    \qquad
    t = - \frac{253}{19} \times (91 \text{~GeV})^{2}
    \text{,}
\end{equation}
and find agreement to the default precision of \texttt{pySecDec} (3--10 digits).

In addition, we also check the self-consistency of the differential equations.
Evaluations at two different phase space points should be connected by a system of differential equations with respect to $s$ and $t$.
We start from the phase point eq.~\eqref{equ:crosscheck_point},
and move to the following two phase space points using differential equations in $s$ and $t$:
\begin{align}
    s_{1} & = \frac{400}{23} \times (91 \text{~GeV})^{2}
    \text{,}
    \qquad
    t_{1} = - \frac{253}{19} \times (91 \text{~GeV})^{2}
    \text{;}
    \\
    s_{2} & = \frac{400}{19} \times (91 \text{~GeV})^{2}
    \text{,}
    \qquad
    t_{2} = - \frac{253}{23} \times (91 \text{~GeV})^{2}
    \text{.}
\end{align}
We compare the results with those obtained by directly solving the $m_{t}^{2}$ differential equation at $(s_{1}, t_{1})$ and $(s_{2}, t_{2})$.
We find that master integrals evaluated in the two different ways agree
up to the precision used when solving the differential equations (20 digits).

\section{Helicity amplitudes}\label{sec:helicityamplitudes}

Having discussed all the preliminary steps in the previous sections, we turn to the evaluation of the helicity amplitudes.
We note that for each phase space point the numerical IBP reduction, evaluation of master integrals, and expansion of the form factors in $\epsilon$ require approximately 5 hours on a single CPU core.

The form factors are insensitive to the polarisation of external particles, which only enters through the tensor structures defined in eq.~\eqref{eq:ampdecomp}. The tensor structures are evaluated by constructing polarisation vectors for the external particles using spinor-helicity formalism. For the gluons the polarisation vectors are given by
\begin{align}
\epsilon_{1,L}^\mu = - \frac{1}{\sqrt{2}} \frac{\spba{2}{1}}{[21]},\quad \epsilon_{1,R}^\mu = \frac{1}{\sqrt{2}} \frac{\spab{2}{1}}{\la 21 \ra}, \\
\epsilon_{2,L}^\mu = - \frac{1}{\sqrt{2}} \frac{\spba{1}{2}}{[12]},\quad \epsilon_{2,R}^\mu = \frac{1}{\sqrt{2}} \frac{\spab{1}{2}}{\la 12 \ra}.
\end{align}

For on-shell production of $Z$ bosons there is a total of $2 \times 2 \times 3 \times 3 = 36$ helicity amplitudes,
but only $8$ of them are independent~\cite{Glover:1988fe,Davies:2020lpf,Agarwal:2020dye}.
For presenting our results, we only consider decays of the $Z$ bosons to two massless fermions.
In this case, the polarisation vectors of the $Z$ bosons can be represented by the following currents in the spinor-helicity formalism,
\begin{align}
\epsilon_{3,L}^{* \mu} = \epsilon_{3,R}^{\mu} = \spab{5}{6},\quad \epsilon_{4,L}^{* \mu} = \epsilon_{4,R}^{\mu} = \spab{7}{8}.
\label{eq:leptopncurrents}
\end{align}
The spinors represent massless fermions in the decays $p_3 \to p_5 + p_6$ and $p_4 \to p_7 + p_8$.
Note that we omit propagators and couplings related to the decays.
The equivalence of conjugation and interchange of leptons $5 \leftrightarrow 6$ and $7 \leftrightarrow 8$ further reduces the number of helicity configurations to two.
We label the two independent helicity amplitudes according to the helicities of particles $1, 2, 5, 7$: $LLLL$ and $LRLL$.

In table~\ref{tab:ampevaluation} we present renormalised helicity amplitudes for the phase space point given in eq.~\eqref{eq:pspoint}.
\begin{equation}
\small{
	\begin{matrix}
	p_{1} & = & ( & 105.0777489925, & 0, & 0, & 105.0777489925 & )
	\\
	p_{2} & = & ( & 105.0777489925, & 0, & 0, & -105.0777489925 & )
	\\
	p_{5} & = & ( & 46.78438395949, & 6.933040148340, & -41.21963371304, & -21.01554979850 & )
	\\
	p_{6} & = & ( & 58.29336503303, & 41.21963371304, & 41.21963371304, & 0 & )
	\\
	p_{7} & = & ( & 71.75825510920, & -71.71311393197, & 0, & -2.544890272092 & )
	\\
	p_{8} & = & ( & 33.31949388331, & 23.56044007059, & 0, & 23.56044007059 & )
	\end{matrix}}\label{eq:pspoint}
\end{equation}
With the masses, $m_{t}$ and $m_{Z}$, fixed, see eq.~\eqref{eq:fixedmasses}, this corresponds to
\begin{align}
s = \frac{132496}{3}\text{~GeV}^2,\qquad t = - \frac{91091}{5}\text{~GeV}^2.
\end{align}

Extrapolating the precision loss order by order in $\epsilon$ in table~\ref{tab:ampevaluation} we find a precision of around $10$ digits for the finite part.
This precision is expected to be attainable for the bulk of partonic phase space.
Closer to the production threshold ($\sqrt{s} = 182$ GeV) we observe strong cancellation and precision drops as a result thereof.
Since the precision of the helicity amplitudes is governed by the master integral evaluations, it can be increased to the desired level by reevaluating the master integrals with more digits.

The partonic phase space can be parametrised using the relative velocity of the $Z$ bosons, $\beta$, and the scattering angle between $p_{1}$ and $p_{3}$, $\theta$, in the centre-of-mass frame.
The the kinematic invariants are related to $\beta$ and $\theta$ as follows,
\begin{equation}
\label{eq:betacostheta}
s = \frac{4 m_{Z}^{2}}{1 - \beta^{2}},
\qquad
t = m_{Z}^{2} - \frac{s}{2} \left(1 - \beta \cos \theta \right).
\end{equation}
Plots for the interference between the finite remainder of the two-loop amplitude eq.~\eqref{eq:irfactorisation} and the leading order amplitude are given in table~\ref{tab:plots}. We define the interference as
\begin{align}
G = \frac{2\,\text{Re} \left[ (A^{(1)})^\star F^{(2)} \right]}{\vert A^{(1)} \vert ^2},\label{eq:interference}
\end{align}
and present the plots separately for each colour factor.
Note that while we use integer values for the particle masses in the amplitudes $A_{vv}^{[X]}$ and $A_{aa}^{[X]}$, see eq.~\eqref{eq:vvaadecomp}, the weak mixing angle is evaluated using physical $Z$ and $W$ masses~\cite{Zyla:2020zbs},
\begin{align}
m_Z = 91.1876\text{~GeV},\quad m_W = 80.379\text{~GeV}.
\end{align}

We have checked that the form factors satisfy required (crossing) symmetry relations from refs.~\cite{vonManteuffel:2015msa,Davies:2020lpf}.
We have also checked that we obtain the same finite remainders by using the four-dimensional tensor structures in eq.~\eqref{eq:ampdecomp} and the $d$-dimensional tensors used in refs.~\cite{vonManteuffel:2015msa,Caola:2015ila,Davies:2020lpf,Agarwal:2020dye}.
Using four-dimensional tensors makes intermediate expressions simpler and requires fewer terms in the $\epsilon$-expansion of the master integrals.

We compared the results of our calculation with those of refs.~\cite{Davies:2020lpf,Agarwal:2020dye} and found good agreement.
For the result presented in ref.~\cite{Davies:2020lpf}, we compared the helicity amplitudes at two high-energy points ($\sqrt{s} = 1290 \text{~GeV},\ \cos \theta = \pm 0.2$).\footnote{These points are well within the radius of convergence of the mass expansion series used in ref.~\cite{Davies:2020lpf}, where Pad\'{e} approximation has been applied to extent the radius of convergence even further.}
We cross-checked $d$-dimensional form factors against the points provided in ref.~\cite{Agarwal:2020dye}, including finite terms kindly provided by the authors.

\section{Conclusions}\label{sec:conclusions}

We described a numerical calculation of the top quark contribution to the two-loop helicity amplitudes of the process $gg \to ZZ$.
The dependence of the results on the top quark mass is accounted for exactly.
We have presented results for a single benchmark point as well as plots for the finite remainders of the helicity amplitudes across the phase space.
The plots are produced from a grid of amplitude evaluations.
For each phase space point integration-by-parts reductions were performed numerically for integer mass values and the master integrals were evaluated using the auxiliary mass flow method.
This allows for efficient evaluation of the amplitude to, essentially, arbitrary numerical precision.

\acknowledgments

We are grateful to Kirill Melnikov for guidance and encouragement throughout this project and helpful suggestions on the manuscript.
We thank the authors of ref.~\cite{Davies:2020lpf} and ref.~\cite{Agarwal:2020dye} for their assistance in cross-checking our results.
This research is partially supported by the Deutsche Forschungsgemeinschaft (DFG, German Research Foundation) under grant 396021762 - TRR 257.
Feynman diagrams in figure~\ref{fig:family},~\ref{fig:region} and~\ref{fig:boundary} were generated using \texttt{FeynArts}~\cite{Hahn:2000kx}.
\texttt{JaxoDraw}~\cite{Binosi:2008ig} was used for the diagrams in table~\ref{tab:colour}.

\begin{landscape}
\centering
\begin{table}
	\begin{center}
		\begin{tabular}{|c|c|c|c|c|}
			\hline \hline
			\multicolumn{2}{|c|}{$\mathbf{C_A}$} & $\epsilon^{-2}$ & $\epsilon^{-1}$ & $\epsilon^{0}$ \\
			\hline \hline
			\multirow{2}{*}{\textbf{LLLL}} & $A^{(2)}/A^{(1)}$ & \small{$1.0000000000008 - 7.6 \cdot 10^{-13} i$} & \small{$0.8304916142577 + 3.229874368770 i$} & \small{$-3.878332328849 - 3.254364077719 i$} \\
			& IR pole & \small{$1.00000000000000$} & \small{$0.8304916142539 + 3.229874368771 i$} & --- \\
			\hline \hline
			\multirow{2}{*}{\textbf{LRLL}} & $A^{(2)}/A^{(1)}$ & \small{$1.0000000000009 - 1.42 \cdot 10^{-12} i$} & \small{$0.2359507533599 + 2.8851548638498 i$} & \small{$1.5709899577479 + 0.2619850649223 i$} \\
			& IR pole & \small{$1.00000000000000$} & \small{$0.2359507533772 + 2.8851548638517 i$} & --- \\
			\hline \hline
		\end{tabular}
	\end{center}

	\begin{center}
		\begin{tabular}{|c|c|c|}
			\hline \hline
			\multicolumn{2}{|c|}{$\mathbf{C_F}$} & $\epsilon^{0}$ \\
			\hline \hline
			\textbf{LLLL} & $A^{(2)}/A^{(1)}$ & \small{$-5.487100965397 + 0.2839759537883 i$} \\
			\hline \hline
			\textbf{LRLL} & $A^{(2)}/A^{(1)}$ & \small{$-4.498637043876 + 0.004984051942i$} \\
			\hline \hline
		\end{tabular}\hspace{1cm}
		\begin{tabular}{|c|c|c|}
			\hline \hline
			\multicolumn{2}{|c|}{$\mathbf{\triangle^{2}}$} & $\epsilon^{0}$ \\
			\hline \hline
			\textbf{LLLL} & $A^{(2)}/A^{(1)}$ & \small{$-0.4472046190541 - 0.00072891867295 i$} \\
			\hline \hline
			\textbf{LRLL} & $A^{(2)}/A^{(1)}$ & \small{$-0.4401391809472 - 0.00018668545746 i$} \\
			\hline \hline
		\end{tabular}
	\end{center}
	\caption{Evaluation of the renormalised two-loop helicity amplitudes, see eq.~\eqref{eq:A2ren}, for the phase space point defined in eq.~\eqref{eq:pspoint}. We normalise by the one-loop amplitude $A^{(1)} \vert_{\epsilon=0}$ and show the infrared pole structure for comparison, see eq.~\eqref{eq:irfactorisation}. The renormalisation scale is set to $\mu = m_Z = 91~\text{GeV}$.}
	\label{tab:ampevaluation}
\end{table}
\end{landscape}

\begin{table}[ht]
	\begin{center}
		\begin{tabular}{|c|c|c|}
			\hline \hline
			\textbf{Amplitude} & \textbf{LLLL} & \textbf{LRLL} \\
			\hline \hline
			& \multirow{9}{0.35\linewidth}{\includegraphics[width=1\linewidth]{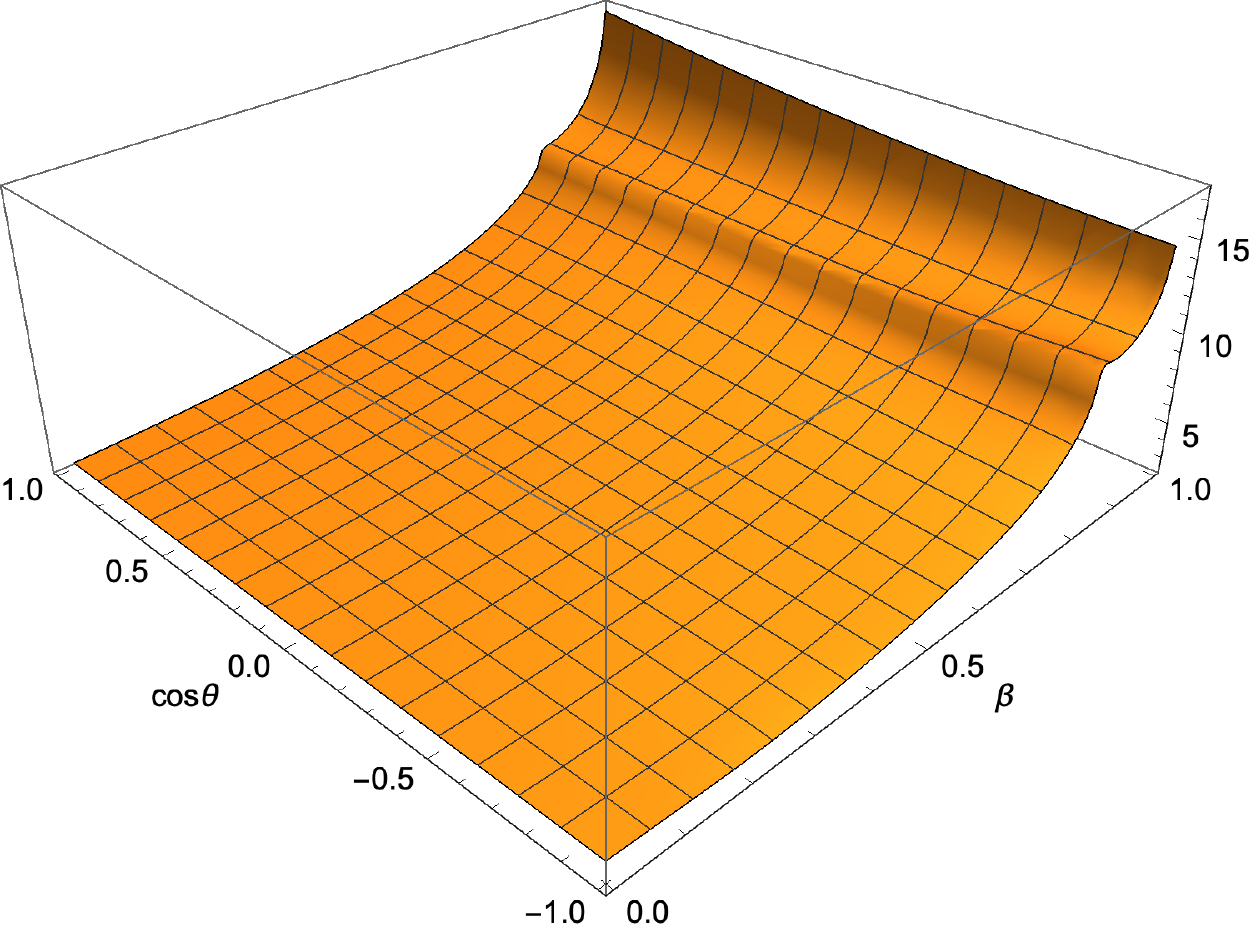}} & \multirow{9}{0.35\linewidth}{\includegraphics[width=1\linewidth]{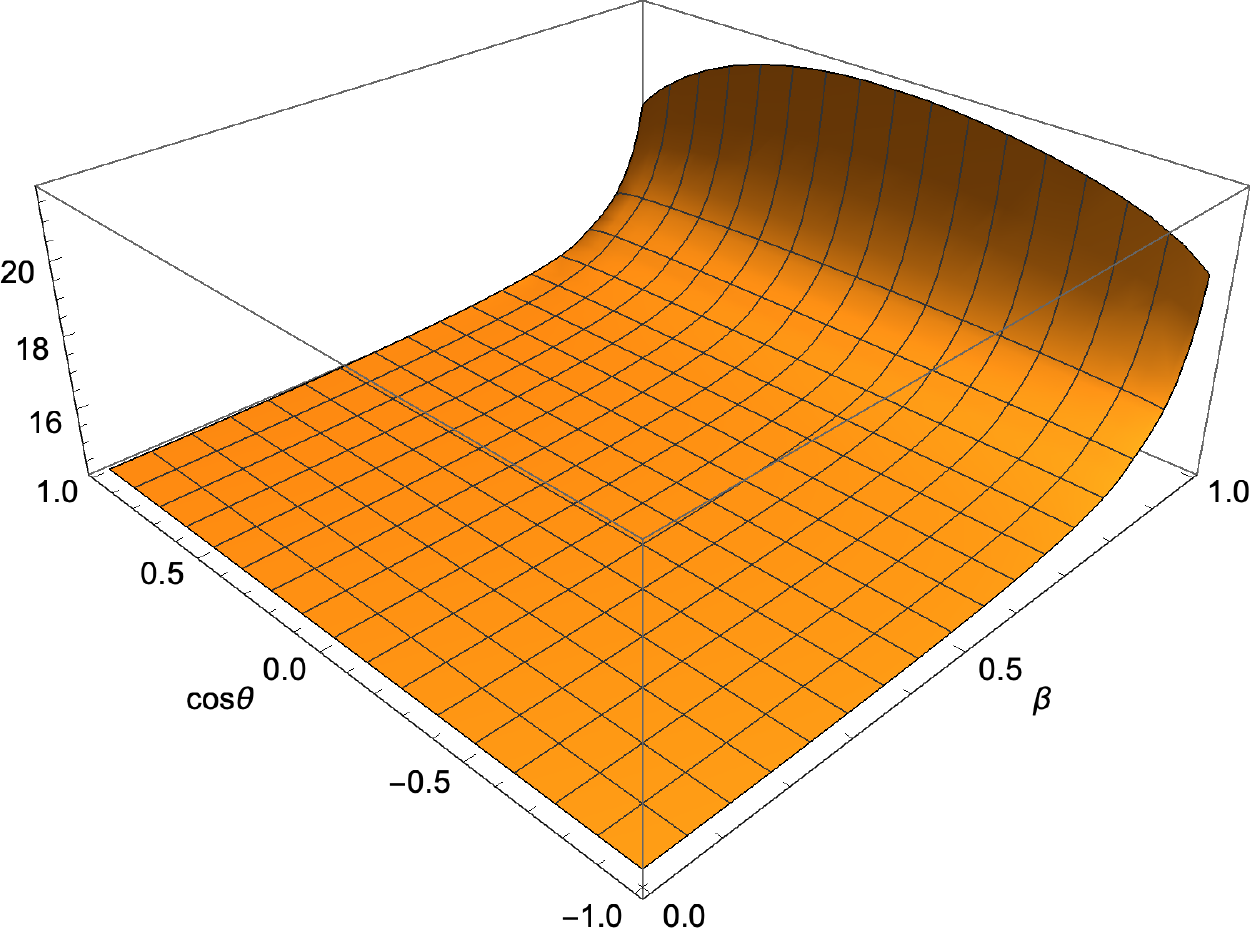}} \\
			& & \\
			& & \\
			& & \\
			$\mathbf{G^{[C_A]}}$ & & \\
			& & \\
			& & \\
			& & \\
			& & \\
			\hline
			& \multirow{9}{0.35\linewidth}{\includegraphics[width=1\linewidth]{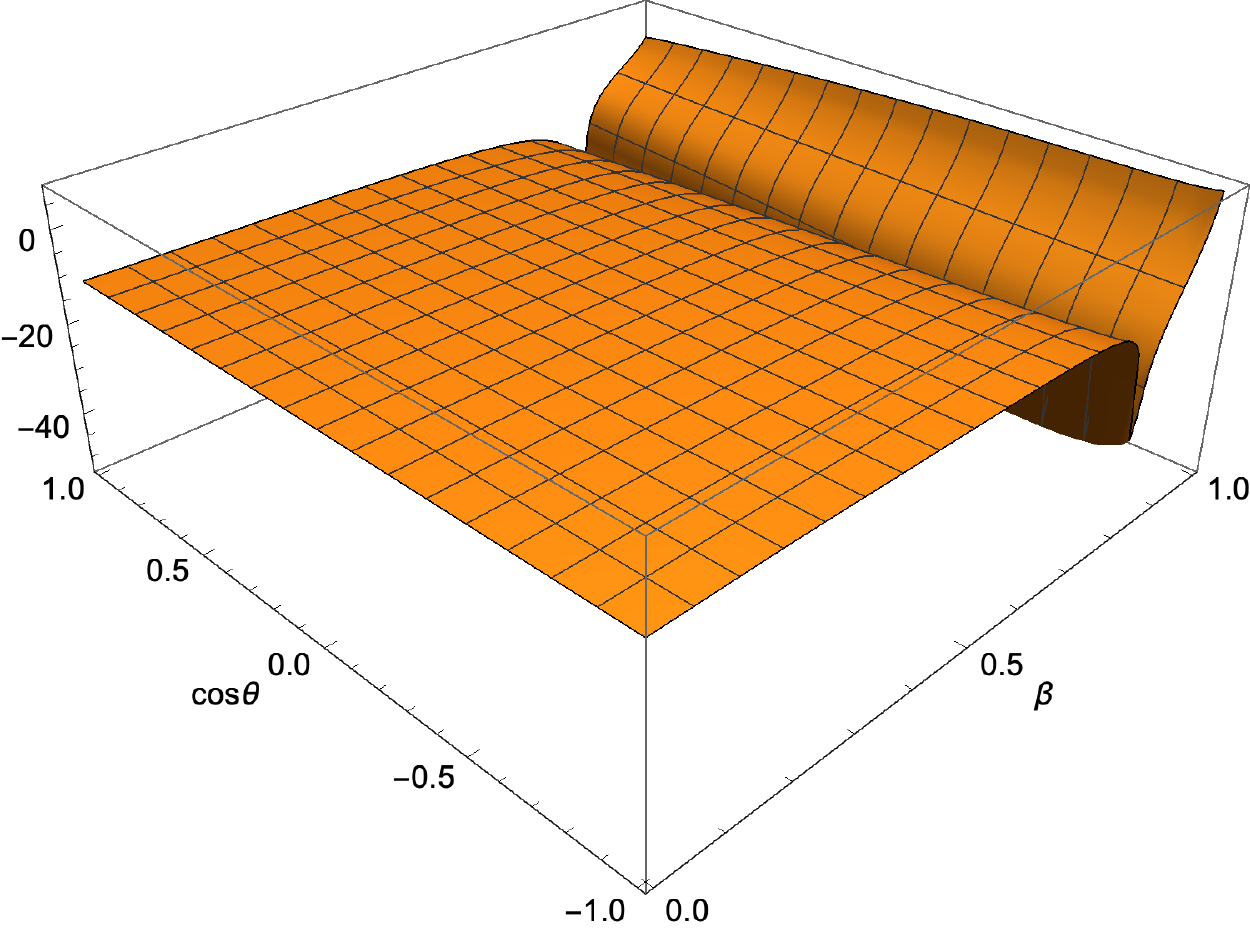}} & \multirow{9}{0.35\linewidth}{\includegraphics[width=1\linewidth]{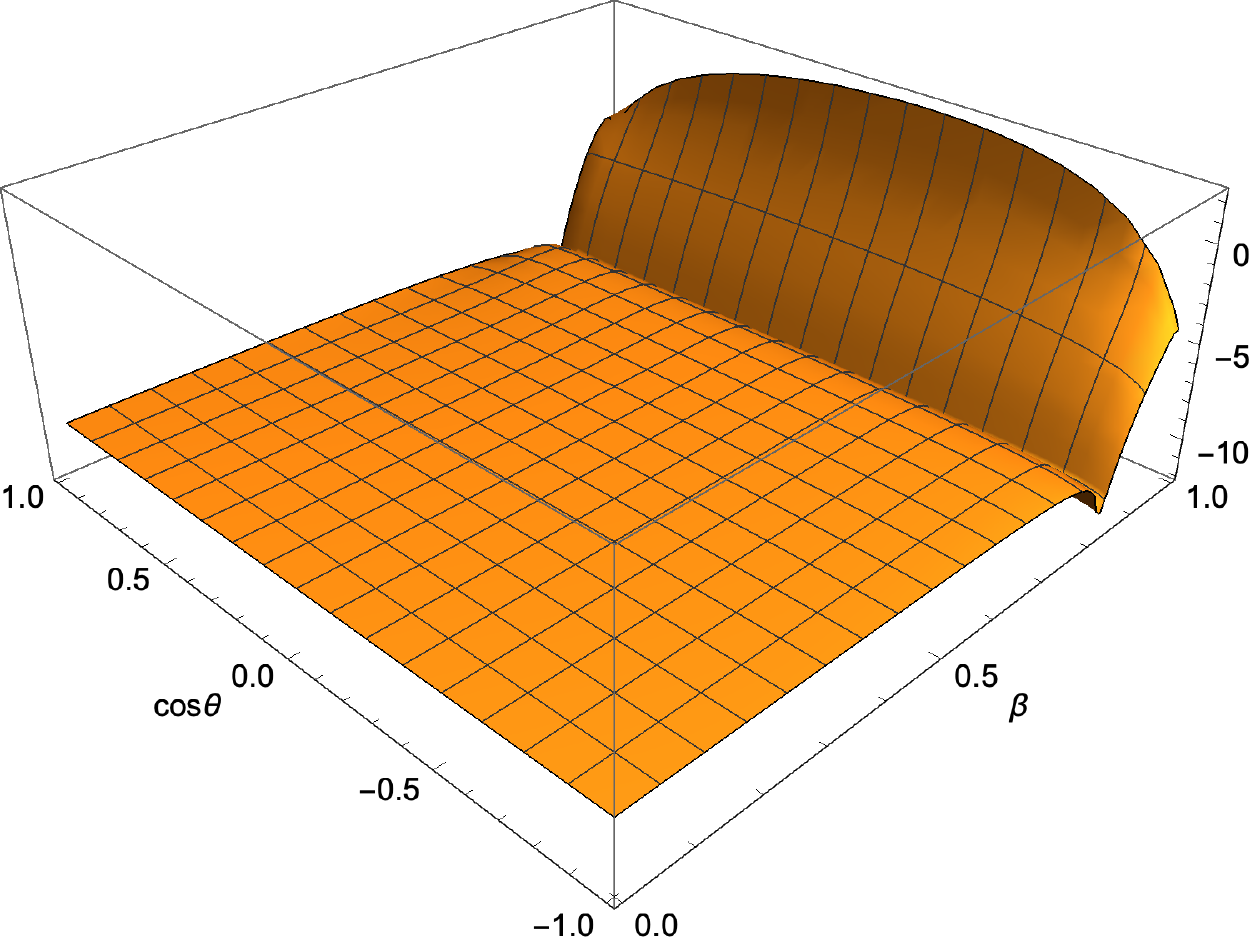}} \\
& & \\
& & \\
& & \\
$\mathbf{G^{[C_F]}}$ & & \\
& & \\
& & \\
& & \\
& & \\
			\hline
			& \multirow{9}{0.35\linewidth}{\includegraphics[width=1\linewidth]{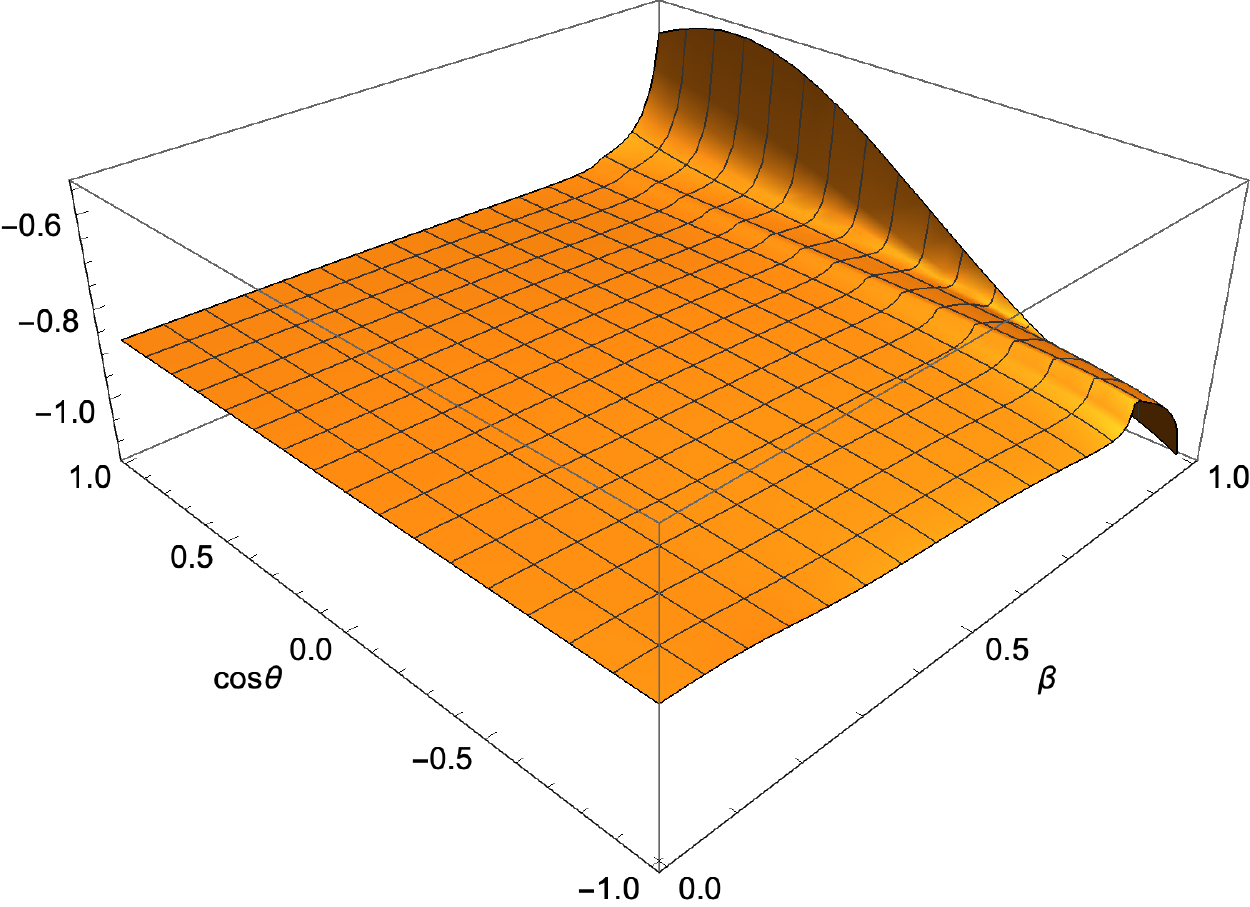}} & \multirow{9}{0.35\linewidth}{\includegraphics[width=1\linewidth]{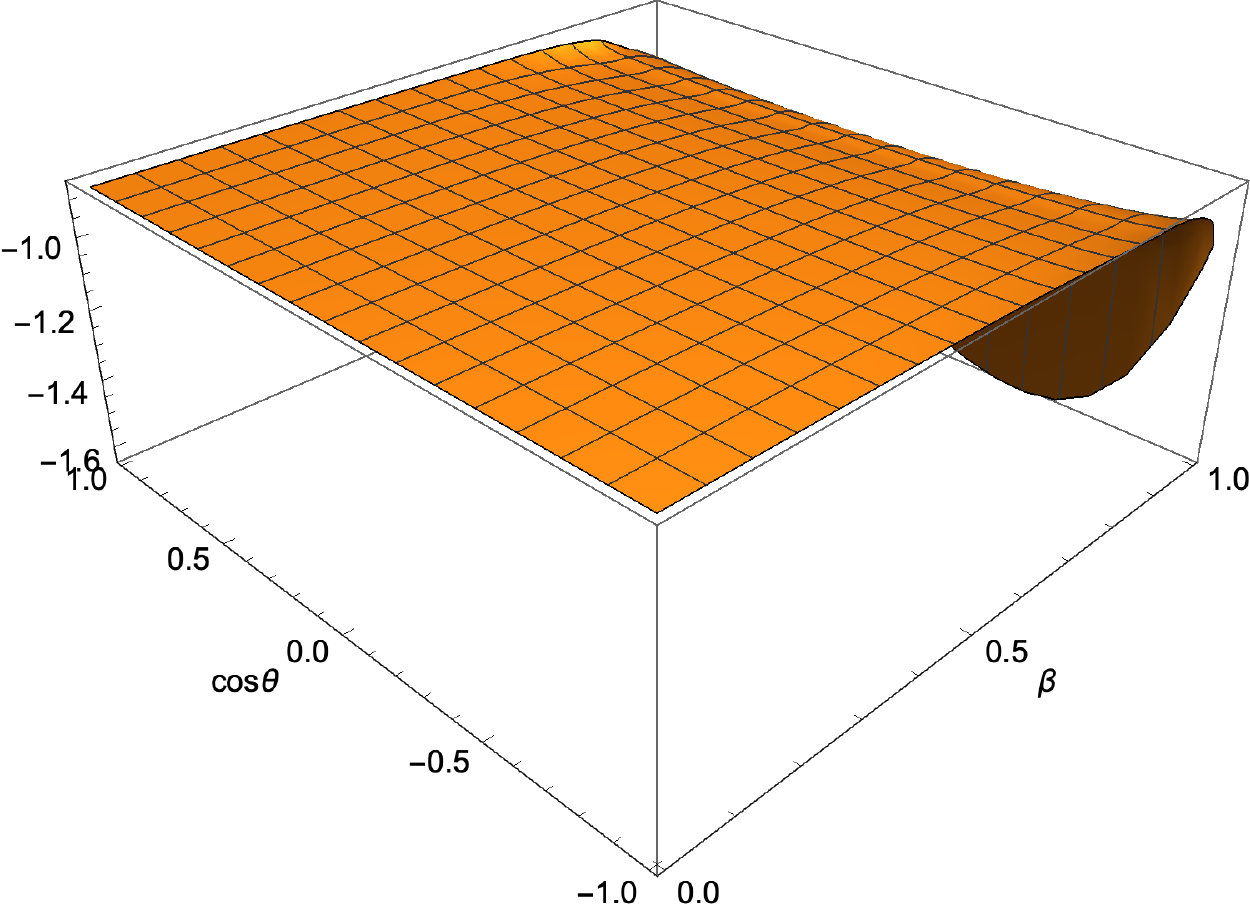}} \\
& & \\
& & \\
& & \\
$\mathbf{G^{[\triangle^2]}}$ & & \\
& & \\
& & \\
& & \\
& & \\
			\hline \hline
		\end{tabular}
	\end{center}
	\caption{Interference of the finite remainders of the two-loop amplitude and the leading order amplitude as functions of $\beta$ and $\cos \theta$. See eqs.~\eqref{eq:betacostheta} and~\eqref{eq:interference} for details. The renormalisation scale is set to $\mu = m_Z = 91~\text{GeV}$.}
	\label{tab:plots}
\end{table}

\newpage

\appendix

\section{Tensor structures}
\label{sec:tensors}

We use the parity even tensor structures of ref.~\cite{Binoth:2006mf} for the amplitude decomposition. They are reproduced here for convenience:
\begin{align}
T_{1} &= \varepsilon_1 \cdot \varepsilon_2\, \eta^{\mu \nu}, \nonumber \\
T_{2} &= \varepsilon_1 \cdot \varepsilon_2\, p_1^\mu p_1^\nu, &
T_{3} &= \varepsilon_1 \cdot \varepsilon_2\, p_1^\mu p_2^\nu, &
T_{4} &= \varepsilon_1 \cdot \varepsilon_2\, p_2^\mu p_1^\nu,&
T_{5} &= \varepsilon_1 \cdot \varepsilon_2\, p_2^\mu p_2^\nu, \nonumber \\
T_{6} &= \varepsilon_2 \cdot p_3 \, \varepsilon_1^\mu p_1^\nu, &
T_{7} &= \varepsilon_2 \cdot p_3 \, \varepsilon_1^\mu p_2^\nu, &
T_{8} &= \varepsilon_2 \cdot p_3 \, p_1^\mu \varepsilon_1^\nu, &
T_{9} &= \varepsilon_2 \cdot p_3 \, p_2^\mu \varepsilon_1^\nu, \nonumber \\
T_{10} &= \varepsilon_1 \cdot p_3 \, \varepsilon_2^\mu p_1^\nu, &
T_{11} &= \varepsilon_1 \cdot p_3 \, \varepsilon_2^\mu p_2^\nu, &
T_{12} &= \varepsilon_1 \cdot p_3 \, p_1^\mu \varepsilon_2^\nu, &
T_{13} &= \varepsilon_1 \cdot p_3 \, p_2^\mu \varepsilon_2^\nu, \nonumber \\
T_{14} &= \varepsilon_1 \cdot p_3 \, \varepsilon_2 \cdot p_3 \, \eta^{\mu \nu}, \nonumber \\
T_{15} &= \varepsilon_1 \cdot p_3 \, \varepsilon_2 \cdot p_3 \, p_1^\mu p_1^\nu, &
T_{16} &= \varepsilon_1 \cdot p_3 \, \varepsilon_2 \cdot p_3 \, p_1^\mu p_2^\nu,\nonumber \\
T_{17} &= \varepsilon_1 \cdot p_3 \, \varepsilon_2 \cdot p_3 \, p_2^\mu p_1^\nu, &
T_{18} &= \varepsilon_1 \cdot p_3 \, \varepsilon_2 \cdot p_3 \, p_2^\mu p_2^\nu.
\end{align}

\section{Integral families}
\label{sec:family}

We define 15 integral families for master integrals.
There are 11 planar families labelled planar 1 to 6, plus crossings ($p_{1} \leftrightarrow p_{2}$) of the first 5 families.
In the nonplanar case,
we define 4 families labelled nonplanar 1 to 3 together with a crossed version of nonplanar 3.
Table~\ref{tab:families} lists their definitions.

\begin{table}[ht]
    \centering
    \begin{tabular}{|c|c|p{0.7 \linewidth}|}
        \hline \hline
        \multicolumn{2}{|c|}{\textbf{Name}} & \multicolumn{1}{|c|}{\textbf{Definition}} \\
        \hline \hline
        \multirow{12}{*}{planar}
        & \multirow{2}{*}{1} &
        $
            l_{1}^{2},
            (l_{1} - p_{1})^{2},
            (l_{1} + p_{2})^{2},
            l_{2}^{2} - m_{t}^{2},
            (l_{2} + p_{3})^{2} - m_{t}^{2},
        $
        \newline
        $
            (l_{1} + l_{2} - p_{1} + p_{3})^{2} - m_{t}^{2},
            (l_{2} - p_{1} - p_{2} + p_{3})^{2} - m_{t}^{2},
            l_{1} \cdot p_{3},
            l_{2} \cdot p_{2}.
        $
        \\
        \cline{2-3}
        & \multirow{2}{*}{2} &
        $
            (l_{1} + l_{2} - p_{1} - p_{2})^{2},
            l_{1}^{2} - m_{t}^{2},
            l_{2}^{2} - m_{t}^{2},
            (l_{1} - p_{1})^{2} - m_{t}^{2},
            (l_{2} - p_{2})^{2} - m_{t}^{2},
        $
        \newline
        $
            (l_{1} - p_{3})^{2} - m_{t}^{2},
            (l_{2} - p_{1} - p_{2} + p_{3})^{2} - m_{t}^{2},
            l_{1} \cdot p_{2},
            l_{2} \cdot p_{3}.
        $
        \\
        \cline{2-3}
        & \multirow{2}{*}{3} &
        $
            (l_{1} + l_{2} + p_{2} - p_{3})^{2},
            l_{1}^{2} - m_{t}^{2},
            l_{2}^{2} - m_{t}^{2},
            (l_{1} - p_{1})^{2} - m_{t}^{2},
            (l_{2} + p_{2})^{2} - m_{t}^{2},
        $
        \newline
        $
            (l_{1} - p_{3})^{2} - m_{t}^{2},
            (l_{2} + p_{1} + p_{2} - p_{3})^{2} - m_{t}^{2},
            l_{1} \cdot p_{2},
            l_{2} \cdot p_{3}.
        $
        \\
        \cline{2-3}
        & \multirow{2}{*}{4} &
        $
            (l_{1} + l_{2} - p_{1} + p_{3})^{2},
            l_{1}^{2} - m_{t}^{2},
            l_{2}^{2} - m_{t}^{2},
            (l_{1} - p_{1})^{2} - m_{t}^{2},
            (l_{1} + p_{2})^{2} - m_{t}^{2},
        $
        \newline
        $
            (l_{2} + p_{3})^{2} - m_{t}^{2},
            (l_{2} - p_{1} - p_{2} + p_{3})^{2} - m_{t}^{2},
            l_{1} \cdot p_{3},
            l_{2} \cdot p_{2}.
        $
        \\
        \cline{2-3}
        & \multirow{2}{*}{5} &
        $
            l_{1}^{2} - m_{t}^{2},
            l_{2}^{2} - m_{t}^{2},
            (l_{1} - p_{1})^{2} - m_{t}^{2},
            (l_{1} + p_{2})^{2} - m_{t}^{2},
            (l_{1} - p_{1} + p_{3})^{2} - m_{t}^{2},
        $
        \newline
        $
            (l_{2} + p_{1} + p_{2} - p_{3})^{2} - m_{t}^{2},
            l_{1} \cdot l_{2},
            l_{2} \cdot p_{2},
            l_{2} \cdot p_{3}.
        $
        \\
        \cline{2-3}
        & \multirow{2}{*}{6} &
        $
            l_{1}^{2} - m_{t}^{2},
            l_{2}^{2} - m_{t}^{2},
            (l_{1} + p_{1})^{2} - m_{t}^{2},
            (l_{1} + p_{3})^{2} - m_{t}^{2},
            (l_{1} - p_{2} + p_{3})^{2} - m_{t}^{2},
        $
        \newline
        $
            (l_{2} + p_{1} + p_{2} - p_{3})^{2} - m_{t}^{2},
            l_{1} \cdot l_{2},
            l_{2} \cdot p_{2},
            l_{2} \cdot p_{3}.
        $
        \\
        \hline \hline
        \multirow{6}{*}{nonplanar}
        & \multirow{2}{*}{1} &
        $
            l_{1}^{2},
            (l_{1} - p_{1})^{2},
            (l_{1} + p_{2})^{2},
            l_{2}^{2} - m_{t}^{2},
            (l_{2} + p_{3})^{2} - m_{t}^{2},
        $
        \newline
        $
            (l_{1} - l_{2} - p_{1})^{2} - m_{t}^{2},
            (l_{1} - l_{2} + p_{2} - p_{3})^{2} - m_{t}^{2},
            l_{2} \cdot p_{1},
            l_{2} \cdot p_{2}.
        $
        \\
        \cline{2-3}
        & \multirow{2}{*}{2} &
        $
            (l_{1} - l_{2} + p_{3})^{2},
            (l_{1} - l_{2} - p_{2} + p_{3})^{2},
            l_{1}^{2} - m_{t}^{2},
            l_{2}^{2} - m_{t}^{2},
            (l_{2} - p_{1})^{2} - m_{t}^{2},
        $
        \newline
        $
            (l_{1} + p_{3})^{2} - m_{t}^{2},
            (l_{1} - p_{1} - p_{2} + p_{3})^{2} - m_{t}^{2},
            l_{2} \cdot p_{2},
            l_{2} \cdot p_{3}.
        $
        \\
        \cline{2-3}
        & \multirow{2}{*}{3} &
        $
            l_{2}^{2},
            (l_{2} - p_{2})^{2},
            l_{1}^{2} - m_{t}^{2},
            (l_{1} - p_{1})^{2} - m_{t}^{2},
            (l_{1} - p_{3})^{2} - m_{t}^{2},
        $
        \newline
        $
            (l_{1} - l_{2} - p_{1})^{2} - m_{t}^{2},
            (l_{1} - l_{2} + p_{2} - p_{3})^{2} - m_{t}^{2},
            l_{2} \cdot p_{1},
            l_{2} \cdot p_{3}.
        $
        \\
        \hline \hline
    \end{tabular}
    \caption{Definitions of the integral families.
        $l_{1}$ and $l_{2}$ are loop momenta while $p_{1}$, $p_{2}$, and $p_{3}$ are external momenta defined in eq.~\eqref{equ:process}.
    }
    \label{tab:families}
\end{table}

\section{Boundary condition of the differential equation}
\label{sec:boundary}

The explicit expressions for the boundary integrals in figure~\ref{fig:boundary} are listed below~\cite{tHooft:1978jhc},
\begin{align}
    I_{1}
    & =
    - \exp(\epsilon \gamma_{E})
    \Gamma(-1 + \epsilon)
    \text{,}
    \\
    I_{2}(q^{2})
    & =
    \exp(\epsilon \gamma_{E})
    \Gamma(\epsilon) (-1)^{\epsilon} (q^{2})^{-\epsilon}
    \frac{\Gamma(1 - \epsilon)^{2}}{\Gamma(2 - 2 \epsilon)}
    \text{,}
\end{align}
where $q^{2}$ corresponds to the four-momentum squared of the external legs.
Note that these boundary integrals are one-loop integrals,
thus they enter the boundary condition through products among themselves.

\bibliographystyle{JHEP}
\bibliography{references}

\end{document}